# Robust Wavelet-Based Watermarking Using Dynamic Strength Factor

Mahsa Kadkhodaei, Shadrokh Samavi

Department of Electrical and Computer Engineering, Isfahan University of Technology
Isfahan, 84156-83111 Iran

*Abstract*— In unsecured network environments, ownership protection of digital contents, such as images, is becoming a growing concern. Different watermarking methods have been proposed to address the copyright protection of digital materials. Watermarking methods are challenged with conflicting parameters of imperceptibility and robustness. While embedding a watermark with a high strength factor increases robustness, it also decreases imperceptibility of the watermark. Thus embedding in visually less sensitive regions, *i.e.*, complex image blocks could satisfy both requirements. This paper presents a new wavelet-based watermarking technique using an adaptive strength factor to tradeoff between watermark transparency and robustness. We measure variations of each image block to adaptively set a strength-factor for embedding the watermark in that block. On the other hand, the decoder uses the selected coefficients to safely extract the watermark through a voting algorithm. The proposed method shows better results in terms of PSNR and BER in comparison to recent methods for attacks, such as Median Filter, Gaussian Filter, and JPEG compression.

*Keywords: watermarking; wavelet transform; copyright protection; adaptive; edge detection.*

## I. Introduction

With the increase in digital media, such as images, videos and voices, in the network environment, protecting the intellectual property of data and securing digital data transfer are becoming essential challenges. As a result, in recent years, copyright protection methods like digital watermarking are major research fields to solve the ownership problem. Watermarking methods generally have two embedding and extracting phases. In the encoder side, copyright information, called watermark, is embedding into contents, and in the decoder side, embedded watermarks are extracting for authentication. Watermarking methods can be classified into different groups like visible and invisible based on imperceptibility with naked eyes. Watermarking methods could also be grouped into blind and non-blind algorithms based on the need for the original image in the extraction phase. We also group the watermarking algorithms into fragile and robust based on their tolerance against accidental or malicious manipulations. Unintentional attacks include JPEG compression, and malicious attacks include the addition of Gaussian noise. Robustness of watermark, which depends on the strength factor, could help a content provider to prove her ownership of the digital media. Robustness and imperceptibility are two opposing qualities of any watermarking method. It worths mentioning that there are attacks that change the watermark in a desired way [1]. In [2] a survey of watermarking approaches, such as compression based, histogram modification based, quantization based, and expansion based in spatial and transform domains, is presented. Frequency domain methods are more robust because the most important visual contents of images are stored in low frequencies. Hence, by embedding the watermark in middle frequencies in a redundant manner, the image remains acceptable to the human visual system, and the watermark will be safe against attacks that do not destroy the whole image. Authors of [3] use visual models to embed a watermark bit according to local image characteristics. They consider frequency sensitivity, luminance sensitivity, and contrast masking as basic features to calculate a just noticeable difference threshold for a transparent, robust, and adaptive embedding in their method. The work presented in [4] has also introduced the complexity of each 8×8 block of the image by computing gradients of blocks. They consider blocks with a high mean gradient as complex, and they embed more than one watermarking bit in such blocks. Their justification is that the human visual system is less sensitive to changes in complex blocks. Reference [5] is a semi-blind method for adaptive watermarking, which is using the entropy of 8×8 blocks. Entropy could determine the texture of images, and thus high entropy blocks in most cases are messy blocks and are good candidates for embedding watermark with high strength factors. The work in [6] applied a flexible and dynamic technique by using a simple linear interpolation equation for watermark embedding in quantized DCT coefficients. The Ridgelet transform is used on high entropy image blocks to embed the watermark in high energy coefficients of the transform [7]. Beside spread spectrum methods, quantization-based methods, watermark into quantized significant features like significant wavelet coefficients. Authors of [8] have quantized the main gradient directions of each level of wavelet transform to embed the watermark into angles of the gradient vector. In [9] normalized correlation between the original image and a random signal is considered as a feature for quantization. Zareian *et al.* [10] have used quantized magnitudes of the low-frequency components to apply a quantization index modulation to the high entropy blocks. There are methods that instead of using intra-block relations, consider inter-block relations too. The algorithm of [11] uses an inter-block coefficient correlation to embed watermarking bits in adjacent blocks adaptively.

In this paper, we propose an adaptive watermarking method in the wavelet transform domain. Our method segments an image into 8×8 blocks and identifies complex blocks and embeds in them with higher strength factors. This means that the strength factor is set dynamically. For blocks with higher complexity, the robustness of the embedded bit is higher than the smooth blocks. Complex blocks are those that contain higher edge pixels. Embedding is done by adding strength factor to the selected coefficients of approximation, horizontal, and vertical wavelet transform of the third level subbands. For the extraction of the watermark, we only compare selected wavelet coefficients with their corresponding coefficients of the original image. Hence, this method can be considered a semi-blind watermarking method. Our results show that the proposed method has higher robustness when compared with recent compatible approaches. Comparisons were performed for attacks such as Median and Gaussian filter and JPEG compression.

The remainder of this paper is as follows. In section II, the proposed embedding algorithm, computation of strength factor, and the extraction algorithm are discussed. In section III, experimental results are shown, and finally, in section IV, we conclude the paper.

## II. PROPOSED METHOD

As Barni *et al.* [12] mention, transformation methods generally have three steps: transforming the image to the desired space, embedding the watermark, and extracting it. In this section, we describe our adaptive watermarking method. The proposed method is embedding the watermark into wavelet transform coefficients of the image. The intended coefficients are those of the third level wavelet of a block. On the other hand, the watermark is extracted by comparing the selected coefficients from the watermarked and the original image. Both parts of the algorithm are explained in the following subsections.

### A. Watermark embedding phase

Figure 3 shows an overview of the embedding of our algorithm. In the embedding phase, the watermark bit is inserted more strongly into blocks that have higher contours or textures than smooth blocks. Complex regions are high-frequency regions, and the human visual system usually ignores changes in such areas. Therefore, stronger embedding in these regions could improve the robustness of our algorithm without corrupting transparency of the watermarked image. Our watermark is a random binary stream of bits. In the first step, the original image is divided into 8×8 blocks, and the complexity of each block is computed as the number of edge pixels. Edges are detected using the canny algorithm. A block which has edge pixels higher than the average number of edge pixels of all image blocks is considered as a complex block. In the next step, the 2-Dimensional Discrete Wavelet Transform (2D-DWT) of each block is computed in three levels using the Haar wavelet. Wavelet transform gives us one approximation matrix named cA, which has low-frequency coefficients. Also, three detail matrices named cH, cV, and cD show higher frequencies in horizontal, vertical, and diagonal orientations at each level. The cA matrix is a lower resolution representation of the original image and could be further decomposed by wavelet transform. Multi-layer decomposition results in a multiresolution pyramid. In this method, we decompose each 8×8 block of the image into three levels, and the coefficient of approximation, horizontal and vertical matrix of the third level are chosen for embedding. If watermark bit $w_i$ is "1", the strength factor ($\alpha$) is added to the chosen coefficients $CA_i$, $CH_i$ and $CV_i$ of block $i$. If $w_i$ is "0," the strength factor is subtracted from the selected coefficients. Hence, $CA_i'$, $CH_i'$, and $CV_i'$ are the embedded coefficients. Equation (1) shows the procedure for the approximation coefficient, and the other two coefficients are computed in the same way.

$$\begin{cases} CA_i' = CA_i + \alpha \ if \ w_i = 1 \\ CA_i' = CA_i - \alpha \ if \ w_i = 0 \end{cases} \quad (1)$$

Because the number of bits of the watermark is normally smaller than the number of blocks of the image, we can embed the watermark more than once in the cover image. Then we can use a voting algorithm for each watermark bit in the extraction phase. Before embedding, the strength factor of a block $i$ is calculated as a function of the standard deviation of approximation subband coefficients of the first level of the wavelet transform. This would show how much this block is complex. In (2) the block complexity is represented as $\sigma_{A_i}$, which has a power of $\gamma$. This is set based on a just noticeable difference (JND) threshold to make the presence of the watermark imperceptible in the image.

$$\alpha = \sigma_{A_i}^{\gamma} \quad (2)$$

The most appropriate $\gamma$ values are tested using five images Lena, Peppers, Boat, Baboon, and Barbara. Attacks such as JPEG compression, salt, and peppers, Gaussian noise, median, and Gaussian filter were performed to find a suitable $\gamma$. The quality of the watermarking process is assessed by Bit-error rate (BER) and peak signal to noise ratio (PSNR). As shown in Figures 1 and 2, a trade-off between these two contradicting measures is necessary to choose the best $\gamma$. We limited this parameter between 0 and 1 because if the value is set to more than 1, PSNR is degraded, and if the value is set to less than 0, the BER of the watermark is unacceptable.

Navneet *et al.* [5] also introduced an adaptive strength factor. Our approach is different than [5], where we consider all of the blocks and the complexity of each image block to change the power $\gamma$. Finally, applying the inverse of the 2-Dimensional Discrete Wavelet Transform gives us the watermarked image.

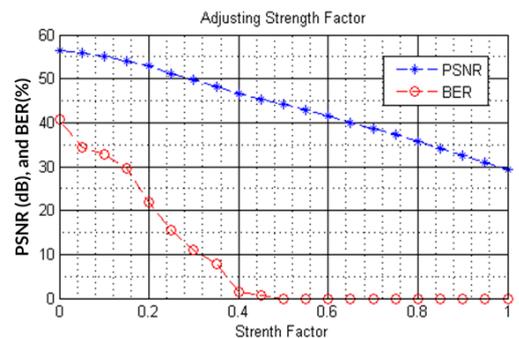

Fig. 1. PSNR and BER of Lena image for adjusting strength factor.

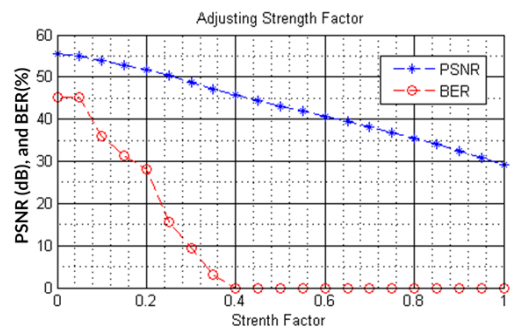

Fig. 2. PSNR and BER of Peppers image for adjusting strength factor.

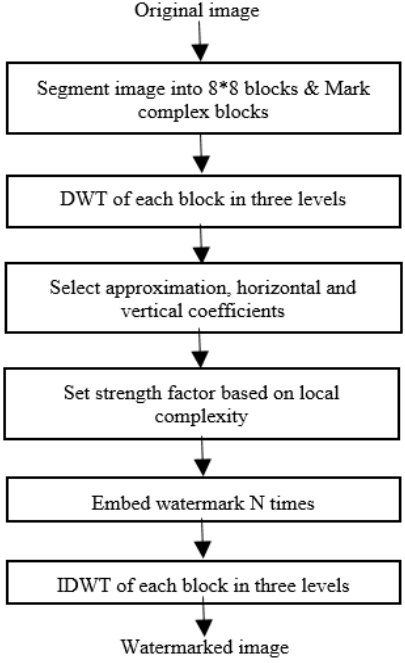

Fig. 4. Embedding part of the proposed algorithm.

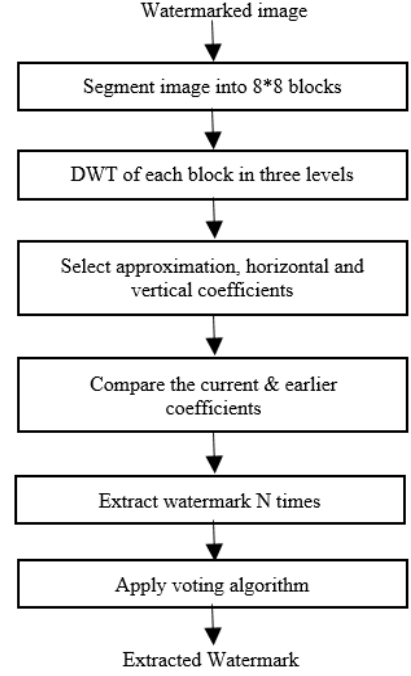

Fig. 3. The extraction part of the proposed watermark.

### B. Watermark extraction phase

Our Extraction schema is represented in Fig. 4. In this phase, the received image is divided into 8*8 blocks, and 2D-DWT is applied in three levels to find the appropriate coefficients of approximation and detail subbands. Now chosen coefficients of this phase and chosen coefficients of the original image in the side information are compared, and if the current coefficient is less than the coefficient before embedding the watermark, we detect watermark bit "1" and if the coefficient before embedding the watermark is less than the current coefficient, we detect "0" bit. Equation (3) shows possibly corrupted selected coefficients of the received image and selected coefficients of the original image as $CA_i''$, $CH_i''$, $CV_i''$ and $CA_i^*$, $CH_i^*$, $CV_i^*$ respectively. Also, $w_i'$ is the extracted watermark from block $i$.

$$\begin{cases} w_i' = 1 \text{ if } CA_i'' > CA_i^* \\ w_i' = 0 \text{ if } CA_i'' < CA_i^* \end{cases} \quad (3)$$

This extraction is done for all of the three coefficients of a block. As in the previous subsection discussed, the watermark can be embedded into the original image more than once. Therefore in this phase, every watermarking bit is extracted more than once, and a voting algorithm can be utilized to select the bit with the majority vote as the final detected bit. This would improve the extraction phase and results in higher robustness.

### III. EXPERIMENTAL RESULTS

In this section performance of our method is evaluated through multiple experiments.

We have assumed that all cover images are of size 512×512 and watermark is a random stream of either 128 or 256 binary bits. The parameter $\gamma$ in (2) is set to 0.4 based on results from numerous experiments with results similar to Fig.1 and Fig. 2. Figure 5 also shows our original and watermarked images for comparison. We show the performance of our algorithm by comparing it with compatible watermarking schemes presented in [5], [7] and [10]. Firstly we compare the proposed method with [10] for attacks such as Median filter, Gaussian filter, salt and peppers, and AWGN. Table I shows the robustness of the proposed method against the median filter, Gaussian filter, salt, and peppers attacks as compared with [10]. In Table II, we are comparing our method with [5] and [7] in terms of BER percent when watermarked images are attacked by the median filter and JPEG compression with different quality factors. All cases that we have better results are shown with **bold** characters. For a fair comparison, we embedded watermarks in such a way that the PSNR of 45 dB was achieved for our algorithm as well as that of [5], [7] and [10]. This robustness is mainly due to the appropriate selection of wavelet coefficients and adaptive selection of strength factors. Results of median filter, Gaussian filter, and salt&pepper attacks are better or comparable with the state of art algorithms.

### IV. CONCLUSION

In this paper, a new adaptive wavelet-based watermarking method was proposed. This algorithm took advantage of image block complexity and human visual system sensitivity in improving the robustness and imperceptibility of the watermark image. We set the strength factor dynamically, based on the block's characteristics, to embed watermark bits in the wavelet coefficients. Appropriate third level wavelet coefficients were selected to enhance the robustness of watermarked images. Our proposed complexity measurement could be used in other applications such as medical image compression [13], or the transmission of images among nodes of visual sensor networks [14]. To prove the functionality of our proposed approach, we compared our experimental results with recent compatible methods, and we showed that our algorithm had a better bit error rate (BER).

TABLE I.  BER (%) OF THE PROPOSED METHOD AND [10] (MESSAGE LENGTH=128 BITS, PSNR=45 dB)

| Attacks | Images | | | |
|---|---|---|---|---|
| | *Lena* | *Boat* | *Peppers* | *Barbara* |
| Proposed Median Filter 3×3 | **0** | **0** | **0** | **0** |
| Ref. [10] Median Filter 3×3 | 1.5625 | 5.4648 | 6.2500 | 4.6875 |
| Proposed AWGN $\sigma^2$=15 | 4.6875 | 0.7813 | 1.5625 | 1.5625 |
| Ref. [10] AWGN $\sigma^2$=15 | 0.0781 | 0.2344 | 0.2344 | 0.5469 |
| Proposed salt & pepper 0.03 | **0** | **0** | **1.5625** | **0** |
| Ref. [10] salt & pepper 0.03 | 7.6563 | 5.6250 | 7.2656 | 5.7813 |
| Proposed salt & pepper 0.04 | **0** | **0** | **3.1250** | **3.1250** |
| Ref. [10] salt & pepper 0.04 | 4.9219 | 7.3438 | 9.4531 | 8.9844 |
| Proposed salt & pepper 0.05 | **4.6875** | **1.5625** | **3.1250** | **1.5625** |
| Ref. [10] salt & pepper 0.05 | 8.5156 | 11.9531 | 13.2813 | 12.1094 |
| Proposed Gaussian Filter $\sigma^2$=1.5 | 0 | **0** | **0** | **0** |
| Ref. [10] Gaussian Filter $\sigma^2$=1.5 | 0 | 1.5625 | 2.3438 | 0 |

TABLE II.  COMPARISON OF THE PROPOSED METHOD WITH [5] AND [7] IN TERMS OF %BER (MESSAGE LENGTH=256 BITS, PSNR=45 dB)

| Attacks | Images | | | |
|---|---|---|---|---|
| | *Boat* | *Baboon* | *Peppers* | *Lena* |
| Proposed Median Filter 3×3 | **0.78** | **0.78** | 0.78 | **0** |
| Ref. [5] Median Filter 3×3 | 1.95 | 7.81 | 0.78 | 0.39 |
| Ref. [7] Median Filter 3×3 | 14.38 | 20.47 | 3.44 | 10.16 |
| Proposed Median Filter 5×5 | 11.71 | **21.09** | 13.28 | 9.37 |
| Ref. [5] Median Filter 5×5 | 9.37 | 24.22 | 5.86 | 1.95 |
| Ref. [7] Median Filter 5×5 | 26.10 | 36.71 | 14.45 | 21.48 |
| Proposed JPEG Q=20 | 3.90 | 0.78 | 3.12 | 7.03 |
| Ref. [5] JPEG Q=20 | 6.25 | 0.39 | 3.91 | 8.98 |
| Ref. [7] JPEG Q=20 | 1.56 | 0.26 | 2.21 | 1.74 |
| Proposed JPEG Q=30 | 0.78 | 0 | **0** | 0.78 |
| Ref. [5] JPEG Q=30 | 1.72 | 0 | 1.17 | 4.30 |
| Ref. [7] JPEG Q=30 | 0.39 | 0.13 | 0.78 | 0.78 |
| Proposed JPEG Q=40 | 0 | 0 | **0** | **0** |
| Ref. [5] JPEG Q=40 | 0.39 | 0 | 0.78 | 0.78 |
| Ref. [7] JPEG Q=40 | 0 | 0 | 0.26 | 0.13 |

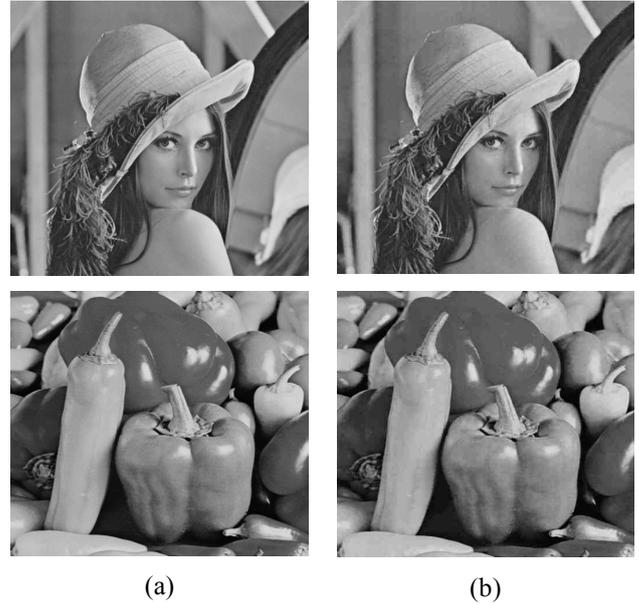

Fig. 5. (a) Original images, (b) watermarked images with PSNR= 45 dB.